# Physics-Based Seismic Hazard and Risk Assessment:

# A New Paradigm for Earthquake Forecasting

**Davide Zaccagnino[1,2], * & Didier Sornette[1], ***

[1] Institute of Risk Analysis, Prediction and Management (Risks-X), Academy for Advanced Interdisciplinary Studies, Southern University of Science and Technology (SUSTech); 1088 Xueyuan Rd., Shenzhen, Guangdong, China, 518055
[2] Istituto Nazionale di Geofisica e Vulcanologia (INGV), Rome, Via di Vigna Murata, 605, 00143, Italy.

**\*** Davide Zaccagnino, zaccagnino@sustech.edu.cn;  Didier Sornette, dsornette@ethz.ch

## Abstract

**Epistemic uncertainty in probabilistic seismic hazard assessment (PSHA) is commonly addressed through a logic-tree framework that combines weighted alternative models to characterize the range of plausible hazard outcomes. Implicit in this approach is a critical assumption: that the available model class provides an adequate representation of the underlying physics governing fault networks. Yet current formulations remain highly simplified, neglecting nonlinear interactions, diverse fault slip modes, multi-scale coupling, and the emergent dynamics that govern the nucleation and evolution of large earthquakes. As a result, the standard treatment of epistemic uncertainty may introduce systematic hazard bias and substantially underestimate forecast uncertainty. To formalize this limitation, we introduce SHARP (Seismic Hazard Assessment and Risks with Physics), a new framework that shifts the focus from selecting among imperfect models to quantifying their collective distance from physical and observational constraints. Central to SHARP is the Model Adequacy Distance (MAD), a quantitative metric of model inadequacy. MAD combines (i) a moment-weighted scoring function scaling with seismic moment to reflect the disproportionate social and economic impact of large events and (ii) compatibility measures derived from geodetic observations and statistical properties of seismicity. We illustrate the approach with an application to the frequency-magnitude distribution of Southern California seismicity. SHARP establishes a rigorous foundation for moving beyond conventional epistemic uncertainty toward a physics-grounded framework for seismic hazard and risk assessment.**

## Key Points

1) PSHA logic trees assume models adequately represent complex fault physics, introducing potential bias.

2) Model Adequacy Distance quantifies how far models lie from physical constraints, revealing hidden uncertainty.

3) SHARP connects empirical forecasting performance and physical constraints governing earthquake dynamics.

## Introduction

Epistemic uncertainty seeks to represent our imperfect knowledge of earthquake dynamics by integrating forecasts from multiple models (Field et al., 2014; Marzocchi et al., 2021). This strategy implicitly assumes the true behavior of fault systems lies within the span of these models, that is, that reality can be adequately represented by some combination of their structures and forecasts.

We argue that this implicit assumption contradicts fundamental earthquake physics. Seismicity is inherently complex, nonlinear, and multi-scale (Ben-Zion, 2008; Sornette, 1991). One of the most widely used and empirically successful statistical frameworks, namely the self-exciting point processes of the ETAS class (Ogata, 1988), captures important aspects of this complexity. In ETAS, the conditional intensity depends on the entire past history of events, generating clustering, cascades, and collective behavior, particularly as the branching ratio approaches criticality. Near this regime, the system exhibits scale-invariant features associated with a branching transition.

However, this apparent complexity arises from a linear superposition of triggering contributions: each event increases the future rate additively and independently of others. While this structure successfully reproduces clustering statistics, it does not incorporate nonlinear stress interactions, threshold dynamics, or state-dependent constitutive evolution. In contrast, physically grounded nonlinear triggering frameworks, such as the Multifractal Stress Activation (MSA) model (Sornette and Ouillon, 2005; Ouillon and Sornette, 2005), embed stress transfer, activation processes, and damage accumulation within intrinsically nonlinear and feedback-driven dynamics.

The distinction is therefore not one of history dependence or cascading, since both are well described by ETAS, but of dynamical structure. In ETAS, large events emerge statistically from branching amplification, whereas in nonlinear physics-based models they may arise from cooperative instabilities rooted in evolving stress fields and material state. Current operational models generally lack explicit internal mechanisms for stress-driven large-scale instability, treating major events statistically rather than as outcomes of a fully coupled physical dynamics (Sornette, 2002).

Indeed, while all models are wrong to some degree because they inherently simplify incompletely understood complexities of nature, some models seem implausible (Stein, 2025). In seismic hazard analysis, models often continue to be used not because they have been rigorously validated but because they have never been formally rejected: validation would require testing their predictions against large earthquakes, yet such events are too infrequent to provide statistically reliable evidence (Mulargia et al., 2017). Addressing this issue requires that we assess not only a model's statistical fit to relatively small events, but its physical plausibility. For instance, hazard maps sometimes assign higher probabilities of large earthquakes to slow deforming continental areas than to rapidly deforming interplate boundaries (e.g., New Madrid seismic zone vs California (Stein, 2025)), exemplifying a model that satisfies statistical criteria yet violates the fundamental mechanical constraint linking deformation to seismicity. As a further examples, Poissonian renewal models, that imply faults deemed overdue are no more likely to rupture, contradict the basic physics of stress accumulation; nevertheless, they remain part of the standard operational toolkit. Moreover, although most models include earthquake triggering, they commonly neglect stress shadowing and treat seismicity as effectively decoupled from crustal deformation, ignoring feedbacks from slow slip and aseismic creep (Bird and Liu, 2007). Many also suffer from the ultraviolet divergence problem inherent in ETAS-like formulations, whereby the predicted hazard becomes dominated by an unphysical proliferation of arbitrary small earthquakes (Helmstetter, 2003; Li et al., 2025).

These examples reveal a common pathology: models are judged primarily by their ability to reproduce earthquake catalogs dominated by small events, rather than by their consistency with the physics governing large-earthquake generation, which is the central objective of seismic hazard assessment. They persist not because they are demonstratively realistic, but because we lack quantitative metrics to measure their distance from physical reality. Therefore, in this context, standard treatments of epistemic uncertainty risk creating an illusion of control: they quantify disagreement among models that may be systematically distant from reality, while offering no measure of how far these models lie from the true dynamics of earthquake generation (Figure 1).

This theoretical article advances a new perspective: rather than pursuing incremental refinements of existing frameworks, we argue for the development a new generation of models grounded in fundamental physics (Jordan, 2006). Such models should explicitly incorporate nonlinearity, geodetic mechanical coupling and the mechanisms by which large-scale organisation emerges from multi-scale clustering (Sornette and Virieux, 1992). We formalize this vision quantitatively and outline a strategy for constructing a physics-based scoring framework tailored to the forecast of large earthquakes.

The Seismic Hazard Assessment and Risks with Physics (SHARP) we introduce here is enabled by the employment of a Model Adequacy Distance (MAD). It reframes the evaluation of seismic hazard models, shifting the central question from "Does it fit the data?" to "Is it physically plausible? Is it consistent with the data? And does it address the objective of quantifying large earthquakes' risks?"

# Quantifying the adequacy of forecasting models

The standard epistemic framework operates on the principle of selecting, from among the available options, the explanation that appears best supported by the data. A common extension in seismology and hazard analysis considers a set of models $M_1, M_2, \ldots, M_N$, assign them weights $\pi_i$ reflecting belief, skill or uncertainty and build an integrated forecast, rather than relying on a single "best model" (Herrmann and Marzocchi, 2023; Selva et al., 2024). Here, the core error is the assumption that the system's true behaviour, R, lies within the subspace spanned by the existing models.

Let us define the model space, $\mathcal{M}$, as the space of all conceivable models. The set $\{M_i\}$ of available models represents a tiny cluster in $\mathcal{M}$. R could be very distant from this cluster. Standard epistemic uncertainty, $U_{epist}$, measures the dispersion $\sigma_{\{M_i\}}$ within the model set $\{M_i\}$ but it does not account for the distance between R and the centroid of our models $\bar{M}$, i.e., to the systematic bias, $b = d(R, \bar{M})$ of the models. When $b \gg \sigma_{\{M_i\}}$, $U_{epist}$ becomes a severe underestimate of our actual error.

We propose to assess a model's adequacy, at least for specific purposes, following the spirit of Sornette et al. (2007) and Marzocchi and Jordan (2014), rather than continuing the routine practice of selecting models primarily on their ability to reproduce small to moderate seismicity. We define this formally as follows.

Let $\mathcal{C} = \{C_1, C_2, \ldots, C_K\}$ be a set of criteria a model must satisfy to be considered minimally adequate for capturing real earthquake dynamics at the scales relevant for forecasting and hazard. These criteria should reflect established physical constraints rather than only statistical fit. To be operational and non-redundant, these criteria must target independent and physically distinct dimensions of earthquake system behavior. Each should isolate a specific structural property of the dynamics and be quantifiable through objective, benchmark-based scoring.

Here, we focus on the theoretical definition of the problem and its proposed solution without discussing how such criteria should be practically defined and implemented; this important issue is discussed in the next sections.

Model adequacy is formalized as a mapping $\Psi: \mathcal{M} \to \mathcal{A}$, where $\mathcal{M}$ denotes the space of candidate models and $\mathcal{A}$ the adequacy space. The latter is defined by $K$ evaluation criteria (with $K = 3$ in our implementation), so that $\mathcal{A} = [0,1]^K$, a $K$-dimensional unit hypercube whose axes correspond to distinct adequacy criteria. For a given model $M \in \mathcal{M}$, the adequacy mapping yields a score vector $\Psi(M) = (\psi_1(M), \psi_2(M), \ldots, \psi_K(M))$, where each component $\psi_j(M) \in [0,1]$ quantifies the degree to which $M$ satisfies criterion $C_j$. A score of 1 denotes full compliance with the benchmark associated with $C_j$, whereas 0 indicates complete inadequacy relative to that criterion. The Ideal Adequate Model (IAM, $M^*$) perfectly satisfies all criteria: $\Psi(M^*) = \mathbf{1} = (1,1,\ldots,1)$.

We define a set of Model Adequacy Distances ($MAD_\gamma(M)$) as

$$MAD_\gamma(M) = d_\gamma(\Psi(M), \mathbf{1}) = \left(\sum_{j=1}^{K} w_j \left(1 - \psi_j(M)\right)^\gamma\right)^{1/\gamma} \qquad (1)$$

where $w_j$ are weights reflecting the relative importance of each criterion and the exponent $\gamma$ can take any real number value corresponding to different ways to combine the criteria in an overall metric. A high MAD value indicates a 'bad' model, i.e., structurally inconsistent with the nature of fault systems. See Figure 1 for a visual representation.

The choice $\gamma = 2$, inspired by Kennedy and O'Hagan (2001), corresponds to a distance related to a standard deviation of the scores across all criteria:

$$MAD_2(M) = d_2(\Psi(M), \mathbf{1}) = \left(\sum_{j=1}^{K} w_j \left(1 - \psi_j(M)\right)^2\right)^{1/2} \qquad (2)$$

The choice $\gamma = 1$ gives a distance corresponding to a weighted average of the scores:

$$MAD_1(M) = d_1(\Psi(M), \mathbf{1}) = \sum_{j=1}^{K} w_j \left(1 - \psi_j(M)\right) = 1 - \sum_{j=1}^{K} w_j \psi_j(M) \qquad (3)$$

The choice $\gamma \to 0$ gives the weighted geometric average of the scores

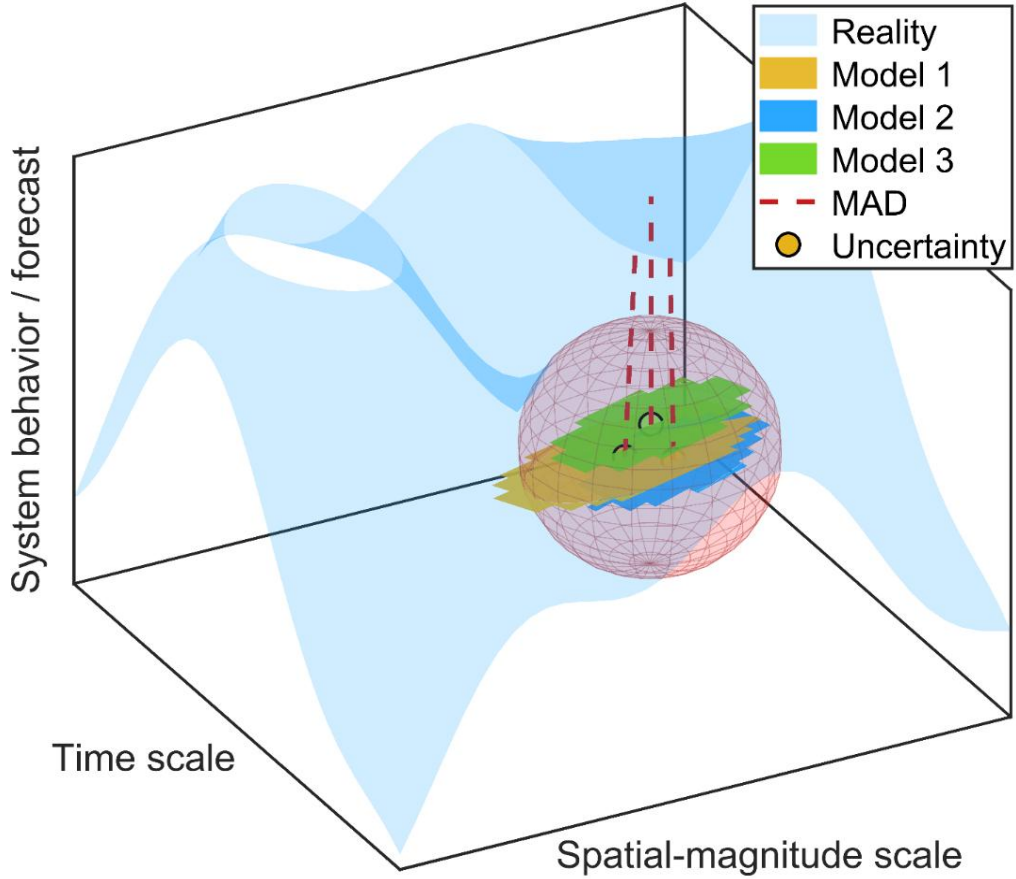

**Figure 1: Conceptual representation of the Model Adequacy Distance (MAD). The true behaviour of seismicity (light blue surface) is complex and extends beyond the limited domains of current models (colored patches). MAD (red dashed lines) quantifies the structural bias of each model, revealing that epistemic uncertainty (pink sphere) merely measures disagreement within an ensemble of assumptions, not the distance to reality.**

$$MAD_0(M) = d_0(\Psi(M), 1) = \prod_{j=1}^{K} \left(1 - \psi_j(M)\right)^{w_j} \tag{4}$$

The choice $\gamma \to +\infty$ gives

$$MAD_{+\infty}(M) = d_{+\infty}(\Psi(M), 1) = Max_{j=1}^{K}\left(1 - \psi_j(M)\right) = 1 - Min_{j=1}^{K}\psi_j(M) \tag{5}$$

and thus considers the criterion that is the least well fulfilled.

The choice $\gamma \to -1$ gives the harmonic mean, which is influenced by the best scores.

The choice $\gamma \to -\infty$ gives

$$MAD_{-\infty}(M) = d_{-\infty}(\Psi(M), 1) = Min_{j=1}^{K}\left(1 - \psi_j(M)\right) = 1 - Max_{j=1}^{K}\psi_j(M) \tag{6}$$

and thus considers the criterion that is best fulfilled.

Summarising, positive $\gamma$ values make the MAD metric dominated by the poorest (worst) adequacy scores, whereas negative $\gamma$ values make it dominated by the strongest (best) scores. For $\gamma > 0$, and increasingly so as $\gamma$ grows, the MAD metric becomes larger whenever any single criterion performs badly; this encourages developing models that perform well in *all* adequacy conditions. For $\gamma < 0$, the metric instead rewards models that excel on one or a few criteria, even if their performance on the others is weaker. The choice of $\gamma$ is informed by the goal of the analysis; we propose, as best practice, to estimate $MAD_\gamma(M)$ as a function of $\gamma$ selected in a range of both positive and negative values in order to get an all-round evaluation for the physical score of each model.

To make the Model Adequacy Distance (MAD) operational for decision making, we reinterpret it as a *trust* variable (Sornette et al., 2007) that directly informs whether a model should be used in scientific, engineering, or societal applications (e.g., for building codes or risk standards). Recall that MAD takes values in [0,1], where $\text{MAD} = 0$ corresponds to perfect adequacy with respect to all empirical constraints, and $\text{MAD} = 1$ to complete inadequacy. We then introduce a mapping

$$\text{trust}: \text{MAD} = m \in [0,1] \to [0,1], \qquad m \mapsto \text{trust}(m), \tag{7}$$

such that $\text{trust}(0) = 1$ and $\text{trust}(1) = 0$, and $\text{trust}(m)$ is strictly decreasing in m.

In analogy with the probability-weighting function of Kahneman and Tversky's Cumulative Prospect Theory (Kahneman and Trevsky, 1979), a standard model of decision-making under uncertainty designed to rank risky alternatives, we interpret $\text{trust}(m)$ as a "subjective" transformation of the "objective" MAD value. For small MAD (nearly adequate models), $\text{trust}(m)$ is convex and lies below the reference line $\text{trust} = 1 - m$, which encodes a rapid loss of trust in response to even modest violations of key constraints. As MAD increases further, the function crosses this reference diagonal and becomes concave, approaching $\text{trust}(1) = 0$ for strongly inadequate models. This shape captures the idea that once a model is already poor, additional degradation in MAD has a comparatively smaller incremental impact on an already low level of trust. The resulting function transforms MAD from a purely descriptive adequacy score into a decision-oriented trust metric that can be directly embedded into engineering safety margins, regulatory thresholds, and model selection procedures. To specify $\text{trust}(m)$, we use a Prelec-type probability-weighting function (Prelec, 1998) to set

$$\text{trust}(m) = \exp[-\eta\,(-\ln(1-m))^\alpha], \quad 0 \le m \le 1; \quad \eta > 0 \text{ and } 0 < \alpha < 1. \tag{8}$$

This parametrisation has the desired properties: $\text{trust}(0) = 1, trust(1) = 0$, it is monotonically decreasing, so that trust decreases as MAD increases; for $0 < \alpha < 1$, the function is convex near m = 0 and concave near m = 1, with a single inflection point. Moreover, for typical values of η and $0 < \alpha < 1$, $\text{trust}(m) < 1 - m$ for small MAD (pessimistic, fast trust loss) and $\text{trust}(m) > 1 - m$ for large MAD, ensuring a single crossing of the line $y = 1 - m$.

See Figure 2 for a visual representation.

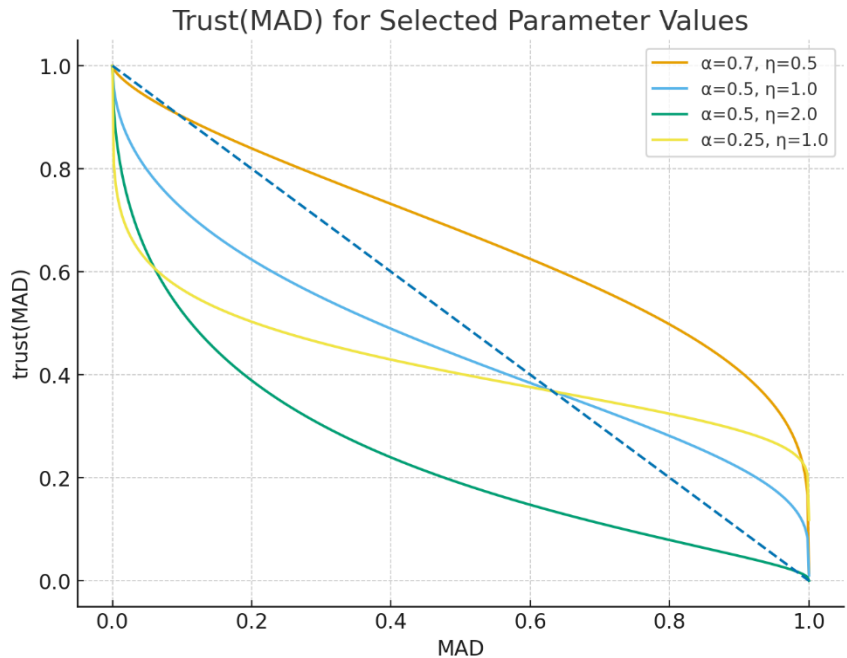

**Figure 2: Trust(MAD) curves for four choices of the Prelec-type parameters $(\alpha, \eta)$. All functions decrease monotonically from $\text{trust}(0) = 1$ to $\text{trust}(1) = 0$, showing rapid early trust loss (convex region) followed by slower decline for highly inadequate models (concave region). The dashed line indicates the linear reference $1 - \text{MAD}$.**

# From adequacy to action: trust, bias correction, and decision-making

The previous section introduced the Model Adequacy Distance (MAD) as a structural measure of how far a model deviates from fundamental physical constraints. A high MAD indicates that a model is structurally inadequate (i.e., it violates the basic physics of earthquake systems) and should be used with caution regardless of its apparent statistical fit. However, measuring inadequacy is only the first step. The question that follows is: how do we translate this structural information into practical hazard assessments?

The standard epistemic hazard is

$$P_{standard}(H) = \sum_{i=1}^{N} \pi_i \cdot P(H \mid M_i) . \tag{9}$$

In this formulation, H denotes the hazardous event or hazard state of interest, e.g., the probability that ground shaking exceeds a threshold, that a structure fails, or that a particular natural hazard scenario occurs. Correspondingly, $P(H)$ is the probability assigned to that hazard event. For each model $M_i$, the quantity $P(H \mid M_i)$ is the conditional probability of the hazard H according to model $M_i$. The weights $\pi_i$ represent the epistemic probabilities assigned to the models themselves in a standard model-ensemble framework. We now proceed to describe how we construct the weights $\pi_i$'s and the conditional probabilities $P(H \mid M_i)$'s for each of the N models.

Given a set $\{M_1, M_2, \ldots, M_n\}$ of available models, we propose to use the trust function defined in the previous section as a weighting mechanism that modulates our confidence in the standard epistemic hazard estimate $P_{\text{standard}}(H)$ defined above. The trust value associated with the ensemble's MAD quantifies the degree to which the aggregated model prediction can be relied upon in decision making. Formally, the weight of model $M_i$ to insert in expression (9) is given by

$$\pi_i = \text{trust}(MAD(M_i)), \tag{10}$$

where $MAD(M_i)$ is the MAD measure of model $M_i$.

To determine the conditional probability $P(H \mid M_i)$ of model $M_i$, we consider a complementary way to quantify structural inadequacy, which is to analyse how each model's performance deteriorates as one considers increasingly large earthquakes. Rather than assuming that past residuals directly approximate future errors, we interpret such residuals as *diagnostics of structural bias*, i.e., empirical indications of how the model diverges from the physical behaviour of seismic systems at different magnitude scales. For a target magnitude threshold $M_{\text{target}}$, we define the magnitude-conditioned residual of model $M_i$ over a hindcast period as

$$\varepsilon_i(M_{\text{target}}) = \frac{1}{N_{\text{obs}}(M_{\text{target}})} \sum_{k: M_k \geq M_{\text{target}}} \left( \log \lambda_i(M_k, x_k, t_k) - \log \lambda_{\text{ref}}(k) \right) \tag{11}$$

where $N_{\text{obs}}(M_{\text{target}})$ is the number of observed earthquakes of magnitude $M \geq M_{\text{target}}$, and $\lambda_{\text{ref}}(k)$ is a non-parametric reference rate (e.g., kernel density or local empirical moment release rate). The associated dispersion, $\sigma_{\varepsilon_i}(M_{\text{target}})$ provides a first empirical estimate of the variability and structural mismatch of model $M_i$ at this scale.

By repeating this analysis over a sequence of increasing $M_{\text{target}}$, we obtain a *magnitude-dependent bias profile* for each model. A systematic increase of $\varepsilon_i(M_{\text{target}})$ with threshold magnitude suggests that the model progressively fails to reproduce the physics of large-event nucleation. Because large events are sparse, this scaling must be estimated over multiple temporal and spatial scales, potentially using spatial multiscale aggregation to stabilize the statistics. In the model-adequacy (MAD) framework, these magnitude-conditioned biases do not define a corrected forecast directly. Instead, they inform a *bias distribution* representing the uncertainty due to structural inadequacy. For a target hazard level H (e.g., the rate of $M \geq 6$ earthquakes), we define a distribution of corrected forecasts (Crow and Shimizu, 1988):

$$P_{\text{adequacy}}(H \mid M_i) = \text{LogNormal}(\log \Phi(M_i) + \mu_{\text{bias}}, \sigma_{\text{bias}}) \tag{12}$$

Depending on the chosen precise hazard variable, $\Phi(M_i)$ may be the expected number of events above a magnitude $M_{\text{target}}$, the annualized rate $\lambda_i(H)$ of the relevant hazard, the intensity-measure parameter used in PSHA (e.g., mean PGA exceedance rate), or any other scalar metric summarizing the model's forecast for hazard H. $\mu_{\text{bias},i}$ and $\sigma_{\text{bias},i}$ are inferred from the sequence $\varepsilon_i(M_{\text{target}})$ and $\sigma_{\varepsilon_i}(M_{\text{target}})$ at a magnitude level consistent with hazard H. This distribution reflects our uncertainty about the structural deviation of model $M_i$ from physical reality.

The log-normal specification in Eq. 12 is motivated by the structural nature of model inadequacy of seismic hazard assessment that we have introduced. First, the true hazard rate is strictly positive; any admissible probability model must respect this constraint. Second, structural bias is inherently multiplicative rather than additive. When a model omits a physical mechanism, such as stress shadowing or fault interaction, the resulting error typically scales with the predicted rate itself. A model that underestimates the occurrence of M6 events by a factor $X$ will generally underestimate M7 events by a similar factor, not by a fixed additive number of events per year. A log-normal distribution naturally reflects this proportional structure: multiplicative deviations in rate become additive deviations in log-space, which are mathematically tractable and statistically interpretable. Third, the log-normal distribution captures the expected asymmetry of structural uncertainty at extreme magnitudes. Underprediction can be arbitrarily large: a model that fails to represent runaway rupture dynamics may underestimate the largest events by orders of magnitude. Overprediction, by contrast, is bounded by the physical constraint that rates cannot fall below zero. The log-normal function accommodates this asymmetry through its heavy right tail, representing the possibility of severe underforecasting.

Alternative distributions are conceivable, each encoding a different view of structural inadequacy. The Gamma distribution, for example, is also positive and skewed but has a lighter right tail than the log-normal function. It would be appropriate only if one assumed that structural bias cannot generate extremely large underestimates, which is an assumption difficult to defend given that current hazard models largely omit the physics governing large-event generation. Fully non-parametric approaches are conceptually appealing but practically infeasible in the large-magnitude regime, where data are sparse and some form of regularized extrapolation is unavoidable.

Then, the adequacy-corrected ensemble hazard becomes

$$P_{\text{adequacy}}(H) = \sum_{i=1}^{N} \pi_i \cdot \text{LogNormal}\left(\log \Phi(M_i) + \mu_{\text{bias},i}, \sigma_{\text{bias},i}\right) \tag{13}$$

which typically exhibits heavier tails than the standard epistemic mixture $P_{\text{standard}}(H)$. This accounts for the empirical fact that models tend to underperform when extrapolated toward the domain of rare high-impact events.

## Operational criteria for assessing models' adequacy

While the previous sections introduced the model adequacy framework from a theoretical perspective, culminating in the general formulation given by Eqs. (9) and (13) with (10), we have not yet specified how it should be implemented in practice. In particular, the construction of the Model Adequacy Distance MAD(M) and the associated trust measure requires explicit definition of the criteria entering the adequacy mapping and the corresponding scoring functions. We therefore now detail how the MAD function is operationally constructed. Specifically, we propose three practical criteria that evaluate a model's ability to: (1) forecast large, high-impact earthquakes through physically consistent mechanisms; (2) remain compatible with independent geodetic observations; and (3) reproduce key statistical properties of seismicity.

**Criterion 1: Moment-Weighted Forecasting Skill (MWFS).** This criterion is the most important because it evaluates whether the model correctly forecasts the rate, location, and timing of large seismic moment release, focusing model evaluation on the societal impact of earthquakes and checking its consistency with the underlying physics of strain energy release. For a testing period T, the model provides a rate density forecast $(\lambda(m, x, t))$. For an observed catalog containing N events $\{(M_i, x_i, t_i), i = 1, \ldots, N\}$, where $M_i$ is the seismic moment-based magnitude, $x_i$ is the epicenter and $t_i$ is the occurrence time of earthquake $i$, the score is built on a Moment-Weighted Log-Likelihood (MW-LL) defined as

$$\text{MWLL} = \sum_{i=1}^{N} w(M_i) \log \lambda(M_i, x_i, t_i) - \int w(m) \lambda(m, x, t) \, dm \, dx \, dt, \tag{14}$$

where the weight $w(M_i) = 10^{3/2\,M_i}$ reflects the seismic moment release. This choice for $w(M_i)$ follows directly from the well-established scaling between moment magnitude and seismic moment, $M_0 \propto 10^{\frac{3}{2}M}$. Since seismic moment $M_0$ is the physically relevant measure of strain energy release, weighting each event by $10^{\frac{3}{2}M_i}$ ensures that the score scales proportionally to the actual seismic moment released by earthquakes. This preserves consistency with energy considerations and prevents the evaluation metric from being dominated by the far more numerous small-magnitude events. This score reconciles model evaluation with the main goal of reproducing the physics of large-event generation. Performance is measured against a null model. The score ($\psi_1$) is defined as a normalized information gain per moment bin using a logistic transformation that maps the real-valued information gain to the interval [0,1]:

$$\psi_1 = \frac{1}{1+\exp(-k\,\mathrm{I})}, \tag{15}$$

with

$$\mathrm{I} = \frac{\mathrm{MWLL}_{\mathrm{model}} - \mathrm{MWLL}_{\mathrm{null}}}{\sum_i w(M_i)}, \tag{16}$$

where $k > 0$ is a sensitivity parameter that controls how rapidly the score increases with positive information gain. This formulation ensures that $\mathrm{I} = 0$ (model performs as well as the null) returns $\psi_1 = 0.5$ while $\mathrm{I} > 0$ gives $0.5 < \psi_1 < 1$, approaching 1 for large positive gains and, finally, that $\mathrm{I} < 0$ for $0 < \psi_1 < 0.5$ as an output, penalizing models worse than the null.

**Criterion 2: Geodetic Strain-Consistency (GSC).** This criterion tests whether the model's long-term seismicity is mechanically consistent with the observed crustal deformation field, ensuring it obeys the fundamental *long-term* tectonic energy budget. For a region ($\Omega$), the long-term seismic moment release rate from the model,

$$\dot{M}_{\mathrm{sim}} = \frac{1}{T_{\mathrm{sim}}} \sum_k 10^{1.5 M_k + 9.1}, \tag{17}$$

where the sum is over events in a sufficiently long simulation ($T_{\mathrm{sim}}$), must equal the tectonic moment accumulation rate, $\dot{M}_{\mathrm{tect}}$, derived from geodetic data (e.g., GNSS velocities)

$$\dot{M}_{\mathrm{tect}} = \mu \int_\Omega \chi(x) \dot{\Sigma}(x)\, dA, \tag{18}$$

where $\mu$ is the average shear modulus, $\chi$ is the seismic coupling which provides a measure of the amount of the long-term geodetic deformation expected to be converted into seismic slip, $\dot{\Sigma}(x)$ is the geodetically inferred strain rate. The score $\psi_2$ evaluates the mechanical consistency of the model by penalizing spatially incoherent or unbalanced moment release. It is defined as

$$\psi_2 = \exp\left(1 - e^{\left(\frac{\dot{M}_{sim} - \dot{M}_{tect}}{\sigma_{\dot{M}}}\right)} + \frac{\dot{M}_{sim} - \dot{M}_{tect}}{\sigma_{\dot{M}}}\right) \times \exp(-D_{KL}(P_{sim} \parallel P_{tect})), \tag{19}$$

and is explicitly composed of two complementary terms. The **first term** measures the consistency of the *total moment budget*: it penalizes discrepancies between the modelled seismic moment release rate $\dot{M}_{\mathrm{sim}}$ and the geodetically inferred tectonic moment accumulation rate $\dot{M}_{\mathrm{tect}}$, normalized by the uncertainty $\sigma_{\dot{M}}$. It is constructed so that the penalty is minimized ($\psi_2 \to 1$) when $\dot{M}_{sim} = \dot{M}_{tect}$, increases approximately linearly when $\dot{M}_{sim} < \dot{M}_{tect}$ (undershoot), and increases exponentially when $\dot{M}_{sim} > \dot{M}_{tect}$ (overshoot). This linear-exponential asymmetry is introduced to encode the physical constraint $\dot{M}_{sim} \lesssim \dot{M}_{tect}$ over long timescales. However, over the relatively short temporal horizons of available data, real overshooting is possible, especially in regions featured by complex tectonic settings. The **second term** evaluates the *spatial coherence* of moment release. It uses the Kullback-Leibler divergence $D_{\mathrm{KL}}$ to quantify the difference between the spatial distribution of simulated seismic moment release $P_{\mathrm{seism}}$ and that of geodetic moment accumulation $P_{\mathrm{tect}}$. This term penalizes models that release moment in locations inconsistent with where strain is accumulating. By construction, $\psi_2 \to 1$ when both the total moment budget and its spatial distribution agree with geodetic constraints and decreases as either the global balance or the spatial pattern becomes inconsistent with tectonic loading.

**Criterion 3: Multiscale Clustering and Emergence (MCE).** This criterion evaluates whether the model spontaneously reproduces the complex, scale-invariant clustering observed in real seismicity, from short-term aftershock cascades to the long-term organization of fault systems. Failure to do so suggests that clustering is imposed through ad hoc prescriptions rather than arising from realistic emergent dynamics. Indeed, natural seismicity displays clustering over a broad continuum of temporal and spatial scales, reflecting the underlying nonlinear processes governing earthquake interactions. As a first-order characterization, a model must reproduce:

(1) Short-term triggering (aftershocks; Utsu and Ogata, 1995). We propose using the branching ratio $n$, i.e., the average number of direct offsprings per event, estimated via the Expectation-Maximization algorithm for an ETAS model fitted to the catalog (Veen and Schoenberg, 2008). This metric is useful to test model's efficiency in generating aftershock cascades. The comparison between the observed branching ratio and the output of the model returns the similarity score $\phi_n = \exp\left(-\frac{|n^{sim}-n^{obs}|}{\sigma_n}\right)$, where $n^{sim}$ is from a long-term simulated catalog and is compared to the observed $n^{obs}$ (with uncertainty $\sigma_n$).

(2) Intermediate- and long-term clustering (Kagan and Jackson, 1991) driven by fault system interactions. The global coefficient of variation given by the ratio between the standard deviation and the mean of interevent times in the catalog, $C_v = \sigma_\tau/\mu_\tau$, is a simple quantity to assess the presence of periods of large-scale activation or quiescence driven by stress transfer beyond direct triggering. As we have already done in the case of the branching ratio and following the same functional form, we introduce the similarity score $\phi_{C_v} = \exp\left(-\frac{|C_v^{sim}-C_v^{obs}|}{\sigma_{C_v}}\right)$.

(3) Multiscale spatial organization reflecting fault network complexity. We propose to employ the fractal correlation dimension $D_2$ (Grassberger and Procaccia, 1983) for the characterization of the spatial patterns of epicenters (or hypocenters, depending on the quality of localizations). This choice is a compromise between computational efficiency and physical accuracy: indeed, seismicity forms multifractal clusters (Geilikman et al., 1990) or even more complex hierarchical structures (Ouillon et al., 1995; 1996) but their spectral reconstruction is time consuming, strongly affected by location uncertainties and often unnecessary, while the fractal dimension can provide a compact measure of the scale-depending clustering of seismicity. Then, using the same approach already employed for the branching ratio and the coefficient of variation, we define the similarity score $\phi_{D_2} = \exp\left(-\frac{|D_2^{sim}-D_2^{obs}|}{\sigma_{D_2}}\right)$.

(4) Scaling of magnitudes as measured by the b-value. While often imposed a priori, a physically consistent model should reproduce the correct Gutenberg-Richter b-value from its internal dynamics. The consistency between simulations and results can be assessed through the similarity score $\phi_b = \exp\left(-\frac{|b^{sim}-b^{obs}|}{\sigma_b}\right)$,

Finally, we quantify the overall ability of the model to reproduce the fractal-scaling-temporal properties of seismicity by combining the four "orthogonal" metrics defined above within the third criterion score defined by

$$\psi_3 = \phi_n \cdot \phi_{C_v} \cdot \phi_{D_2} \cdot \phi_b. \tag{20}$$

The third criterion score is thus defined as the product of the four individual performance measures, so that overall adequacy requires simultaneous success on all components. Because the aggregation is multiplicative, the combined score remains close to 1 only if each individual score is itself close to 1. Even if only one component is small, indicating failure on a single essential criterion, the overall score is proportionally reduced, regardless of strong performance on the others. This multiplicative structure therefore enforces complementarity rather than compensability: deficiencies cannot be "healed" by excellence elsewhere. The product formulation reflects the principle that physical plausibility requires concurrent satisfaction of all core constraints.

The three criteria described above are complementary by construction, jointly spanning distinct and non-overlapping aspects of earthquake dynamics: C1 (MWFS) probes specific predictive skill focused on large events in retrospective tests; C2 (GSC) checks large-scale physical consistency with tectonic forcing; C3 (MCE) assesses the multi-scale organizational structure and emergent properties generated by the model's internal dynamics. Therefore, a model scoring highly on all three would demonstrate its ability to correctly forecast major earthquakes, would operate within the correct energy budget dictated by plate tectonics, and would spontaneously generate the complex spatiotemporal patterns and correct magnitude scaling characteristic of real fault systems. Such a model can be considered "adequate".

It is important to note that the model-adequacy framework does not explicitly penalize model complexity. In contrast to criteria such as AIC or BIC, which introduce theoretical corrections to discourage overparameterization, MAD evaluates model performance empirically rather than through parametric penalties.

MAD assigns greater weight to the largest seismic events. As a result, a model that attains an excellent retrospective fit by introducing additional parameters to capture random fluctuations or transient patterns will inevitably be exposed in prospective testing: subsequent large events will occur at different magnitudes, locations, or under conditions that invalidate those fitted idiosyncrasies. Such failure is not merely a statistical artifact but an empirical indication that the model's added parameters encode assumptions lacking generality.

MAD quantifies this empirical inadequacy, and the associated "trust" transformation translates it into an operational decision rule: models that fail to anticipate the future distribution of large earthquakes, however well they reproduce past data, are structurally inadequate and should be rejected.

## Application to the frequency-magnitude distribution in Southern California

We implement the MAD framework to evaluate six competing earthquake recurrence models in Southern California. The analysis uses seismic events from the SCEC catalog (Hauksson et al., 2012) above the completeness magnitude (M ≥ 2.7). The catalog is temporally divided into a learning period (1 January 1981 to 31 December 2009) for model calibration, and a testing period (1 January 2010 to 31 December 2025) for pseudoprospective evaluation. The testing period is pseudoprospective in the sense that it follows the learning period chronologically and is not used in any model fitting, mimicking a true forward forecast. We have preventively applied our framework on synthetics where catalogs are simulated with the known six different models of frequency-magnitude distribution to check its functioning.

In order to illustrate the MAD framework, we focus on the first and most important criterion, the Moment-Weighted Forecasting Skill (MWFS). Since only one criterion is considered, the definition $MAD_\gamma(M) = d_\gamma(\Psi(M), 1) = 1 - \psi(M)$ becomes independent of the exponent $\gamma$. The analysis is conducted in two different configurations: a whole-region analysis treating the study area as a single, fully connected domain, and a spatially tessellated analysis dividing the region into a 12 × 12 grid (144 squared cells). This dual approach allows us to assess both global model performance and spatial heterogeneity in model adequacy, with implications for understanding the physical processes governing earthquake occurrence across different tectonic settings.

For each spatial cell containing sufficient data, we employ all six models described below using only learning-period data. The null model serves as the reference for computing information gain. We then evaluate each model's forecast against the testing-period data using a moment-weighted log-likelihood score $\Psi_1$ that emphasizes performance on larger-magnitude events. The scores obtained in each cell are aggregated to compute global information gain $I$, from which we derive MAD, and trust values.

### Whole-region analysis

We apply the MAD framework to a set of six competing models for the frequency-magnitude distribution of earthquakes in Southern California, as a first step, to the entire catalog as a single spatial unit. The six models are 1) the standard Gutenberg-Richter (GR) model (Gutenberg and Richter, 1944), 2) the Tapered GR model (Kagan, 2002; Sornette and Sornette, 1999), 3) the Double GR model (Vere-Jones, 2005; Saichev and Sornette, 2005), 4) the GR with extreme event (GR-E) modelled using a Gaussian additional term - this model is statistically equivalent to the proposed characteristic earthquakes (Schwartz and Coppersmith, 1984) or to the dragon-king events (Sornette, 2009), 5) the tapered GR with Gaussian extreme and 6) the Double GR law with Gaussian extreme.

They are described hereafter together with their performance in terms of MAD and trust.
1) We employ the Gutenberg-Richter (GR) model as null hypothesis for reference
$$\lambda(m) = \beta e^{-\beta(m-M_{min})} \tag{21}$$
The Tapered Gutenberg-Richter (GR) model,
$$\lambda(m) = \beta e^{-\beta(m-M_{min})} \cdot \exp\left(-\frac{10^{b(M_x-m)}}{b \ln 10}\right), \tag{22}$$

where $M_x$ is the Kagan's corner magnitude, $\beta = b\ln(10)$, and b is the b-value,
achieved the highest performance of I = 1.276, $\psi_1$ = 1.00, and almost top trust = 0.99 (Figure 3). Its excellent performance indicates that the progressive deficit of very large earthquakes relative to unbounded GR extrapolation is a stable, generalizable property of California seismicity at large scales.

The Double GR model,

$$\lambda(m) = \begin{cases} \beta_s e^{-\beta_s(m-M_{min})}, m < M_p \\ \beta_l e^{-\beta_l(m-M_p)}, m \geq M_p \end{cases} \qquad (23)$$

with the constraint $b_l > b_s$, also performs well (I = 0.140, $\psi_1$ = 0.802, trust = 0.625, Figure 3). The positive information gain confirms that the magnitude-frequency slope steepens at larger sizes over a quite large range of magnitudes. This is, like in the previous case, a physical effect of finite fault dimensions. However, the lower trust value reflects instability in the estimated pivot size $M_p$ within the magnitude range, penalizing this additional complexity relative to the smoother taper formulation.

The GR with extreme earthquake (GR-E) model, defined as

$$\lambda(m) = \Lambda_{bg}\beta e^{-\beta(m-M_{min})} + \Lambda_{ext}\frac{1}{\sigma\sqrt{2\pi}}e^{-\frac{(m-M_{ext})^2}{2\sigma^2}}, \qquad (24)$$

performs slightly below the null (I = -0.01, $\psi_1$ = 0.48, trust = 0.42, Figure 3). Here $\Lambda_{bg}$ and $\Lambda_{ext}$ are weight coefficients and $\sigma$ provides the measure of the width of the Gaussian contribution generated by the extreme event. Physically, this indicates that extreme earthquake behavior, defined as repeat ruptures of nearly identical size, is not a stationary property of fault systems at the spatial scales of our catalog compared to the observation period. The specific magnitude of the largest event in the learning period did not predict the largest event in the testing period, and imposing a fixed bump degrades forecast skill.

Tapered GR model with Extreme event reaches a very similar performance compared to the Tapered GR model.

Conversely, the Double GR + Extreme model fails spectacularly in pseudoprospective testing (I = -0.78, $\psi_1$ = 0.00, trust = 0.06, Figure 3). Despite excellent in-sample fit, its highly parameterized structure actually overfits the specific realization of large events in the learning period. The apparent b-value transition and extreme magnitude are not stable properties of the fault system, but, rather, likely artifacts of specific temporal windows. We think this is an excellent example of the MAD principle: structural inadequacy is revealed not by in-sample statistics but by prospective forecasting failure weighted by seismic moment. See Figure 3 for a visual summary.

## Spatially tessellated analysis

Partitioning the study region into T = 144 (12x12) independent cells $c$ introduces spatial heterogeneity that alters model rankings. Here, the total moment-weighted log-likelihood is aggregated from all cells

$$MWLL_{total} = \sum_{c=1}^{T} MW\,LL_c. \qquad (25)$$

The aggregation of log-likelihoods from single cells follows from the principle that the total log-likelihood of spatially independent observations is the sum of their log-likelihoods. Then, the information gain reads

$$I = \frac{MWLL_{model} - MWLL_{null}}{\sum w(M_i)}, \qquad (26)$$

where $w(M_i) = 10^{1.5M_i}$, computed from the summed likelihoods. This aggregation properly weights each cell's contribution by its observed seismic moment release.

The most notable change is the Double GR + Extreme model, which soared from worst to best performer (I = 2.97, $\psi_1$ = 1.00, trust = 1.00, Figure 4). This inversion reveals that its adequacy is highly spatially variable. In cells containing major faults with clear extreme earthquakes (see Figure 5), the model well captures the emergence of large earthquakes beyond the standard predictions of the GR law. In other cells, it overfits noise.

The Double GR and Tapered GR models maintained positive information gain (I = 0.052 and I = 0.063, respectively) with moderate trust values (0.50 and 0.52). Their performance degrades relative to the whole-region analysis because spatial heterogeneity was previously averaged out. The Tapered GR + Extreme shows a modest positive skill (I = 0.02, $\psi_1$ = 0.55, trust = 0.46), while the GR + Extreme model again performed poorly (I = -0.42, $\psi_1$ = 0.015, trust = 0.13), confirming that the simple bump is generally not a well-defined structure in earthquake magnitude statistics.

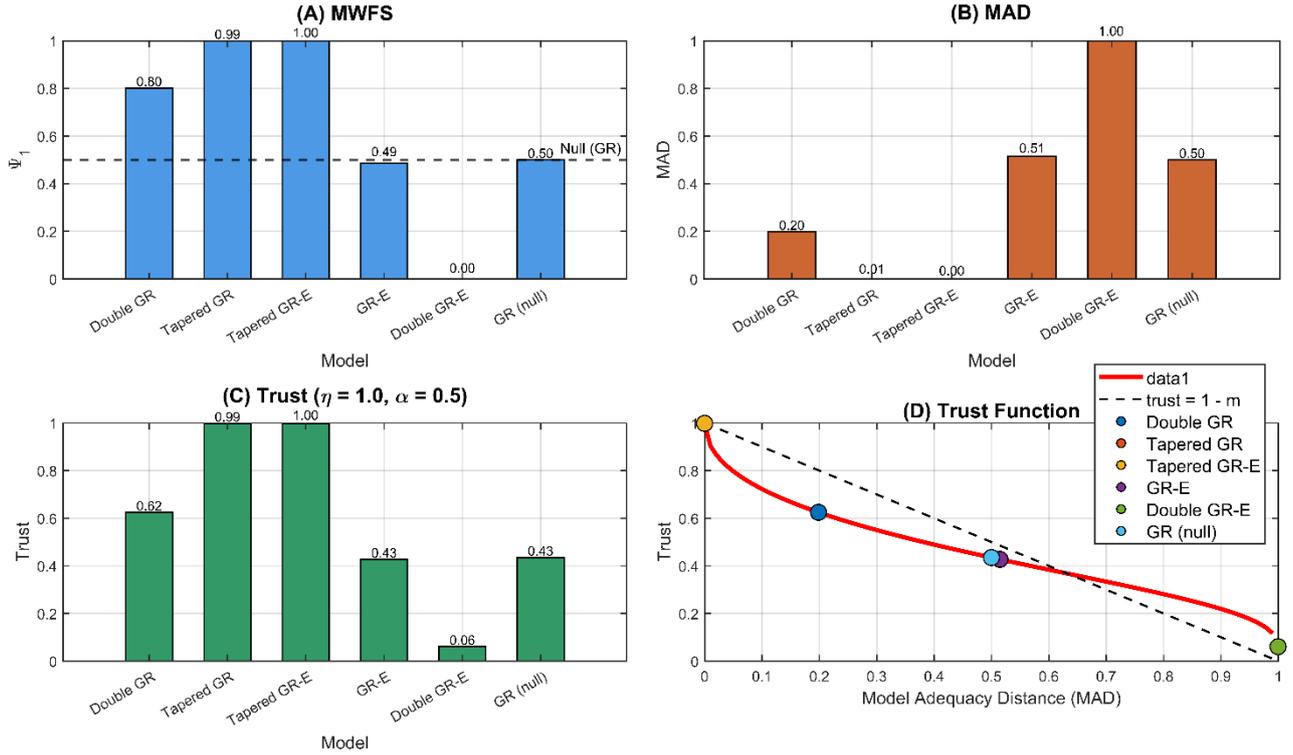

**Figure 3:** Output of six models' assessment for frequency-magnitude distributions in Southern California (seismic events from the SCEC catalog with M ≥ 3 are used - which guarantees completeness; data from 1/1/1981 to 31/12/2009 are employed for calibration, while from 1/1/2010 to 31/12/2025 for testing). The performance of double Gutenberg-Richter (double GR), Kagan's tapered GR, Kagan's tapered GR with extreme earthquake, GR with extreme magnitude earthquake (GR-E) and double GR with extreme earthquake is assessed compared to the standard GR law implemented as null hypothesis. In (A) the output of the Moment-Weighted Forecasting Skill (MWFS, criterion 1) is shown; (B) represents the Model Adequacy Distance (MAD) (since we only consider one criterion, all the definition of MAD given in the main text are equivalent); in (C) the values of trust(MAD) are plotted as bars, while in (D) they are located along the trust curve (selected parameters: $\alpha = 0.5$, $\eta = 1.0$).

The null GR model has $\psi_1$ = 0.50, MAD = 0.50, trust = 0.43 by construction; it can be interpreted as a threshold. Models with trust > 0.43 demonstrate better predictive skills; models with trust < 0.43 are inadequate and should rejected for hazard applications.

## Conclusions

We have developed a physics-based framework (SHARP) for evaluating seismic hazard models that moves beyond conventional measures of epistemic uncertainty. By introducing the Model Adequacy Distance (MAD), inspired by Kennedy and O'Hagan (2001), we have provided a quantitative metric for assessing structural adequacy rather than merely comparing models within an existing ensemble. The associated trust(MAD) function translates this adequacy into an operational quantity suitable for decision-making contexts.

To ensure that evaluation aligns with the fundamental objective of hazard assessment, we complemented MAD with magnitude-dependent bias diagnostics and a moment-weighted scoring framework. These tools explicitly prioritize the accurate representation of large-earthquake generation and strain-energy release, shifting emphasis away from goodness-of-fit metrics dominated by small events. We illustrated the approach through an application to Southern California seismicity, examining six alternative frequency-magnitude formulations and demonstrating how the framework discriminates among them based on physical plausibility rather than statistical fit alone.

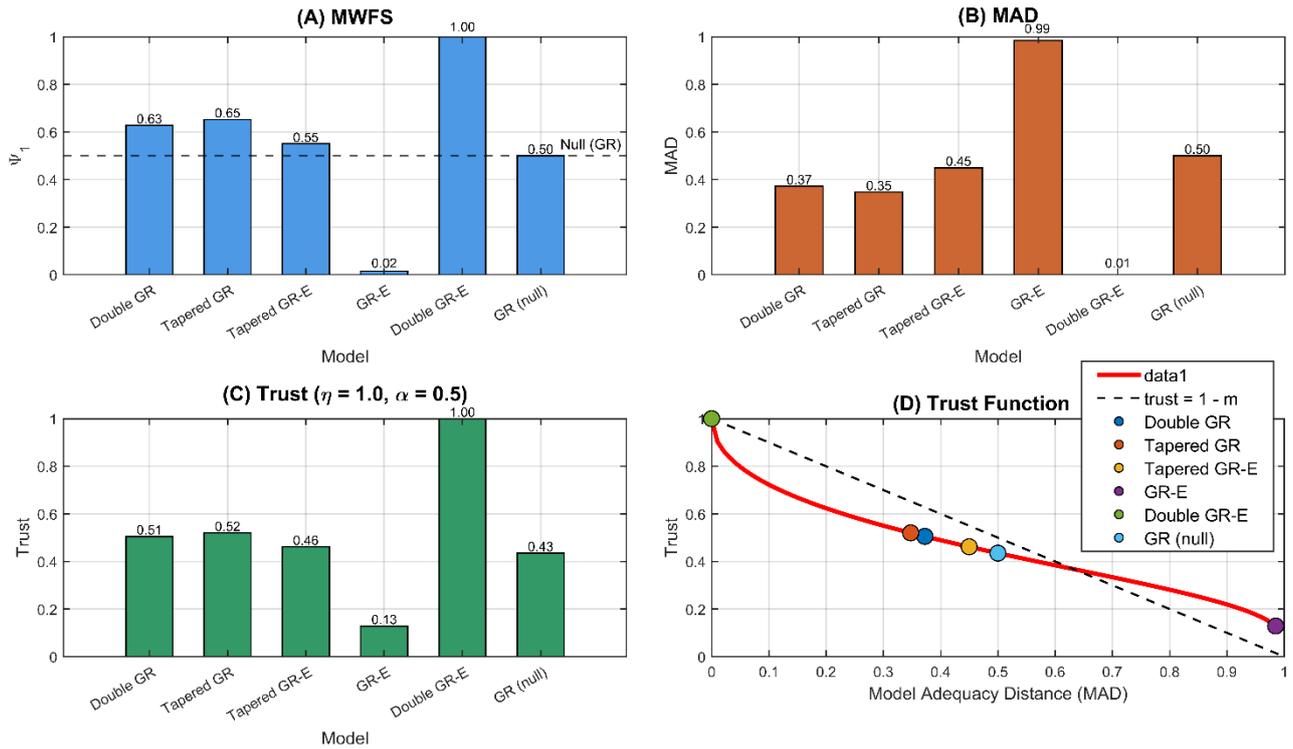

Figure 4: Output of six models' assessment for frequency-magnitude distributions in Southern California (seismic events from the SCEC catalog with M ≥ 2.7 are used - which guarantees completeness; data from 1/1/1981 to 31/12/2009 are employed for calibration, while from 1/1/2010 to 31/12/2025 for testing). Here, the catalog is analysed in its spatial components introducing a grid of 12x12 squared cells; the performance of each model is assessed within each element of the lattice and then aggregated. The performance of double Gutenberg-Richter (double GR), Kagan's tapered GR, Kagan's tapered GR with extreme earthquake, GR with extreme earthquake (GR-E) and double GR with extreme magnitude earthquake is assessed compared to the standard GR law implemented as null hypothesis. In (A) the output of the Moment-Weighted Forecasting Skill (MWFS, criterion 1) is shown; (B) represents the Model Adequacy Distance (MAD) (since we only consider one criterion, all the definition of MAD given in the main text are equivalent); in (C) the values of trust(MAD) are plotted as bars, while in (D) they are located along the trust curve (selected parameters: $\alpha = 0.5$, $\eta = 1.0$).

Together, these developments establish a coherent bridge between empirical forecasting performance and the physical constraints governing earthquake dynamics. Progress in forecasting large earthquakes will not emerge solely from the "data deluge" alone. It requires fundamentally new ways of evaluating, weighting, and constructing models and new conceptual eyes capable of recognising the hidden physics of large-event nucleation and cascade emergence.

## Data and Resources

No data and software were produced to realize this research.

## Acknowledgements

The authors acknowledge useful discussions with Michele Matteo Cosimo Carafa, Carlo Doglioni, Giuseppe Petrillo and Seth Stein.

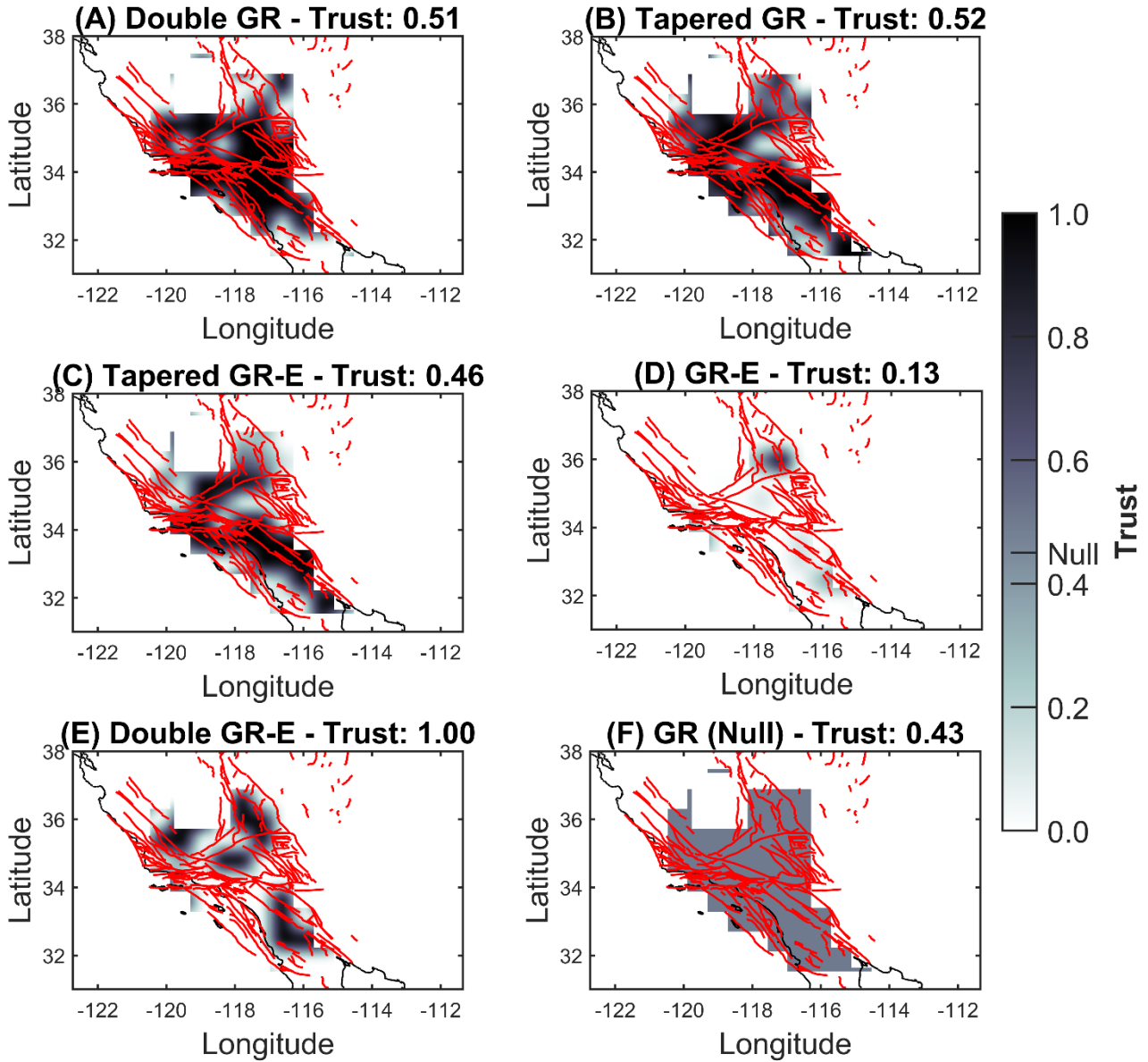

Figure 5: Maps of the spatial distribution of trust(MAD) - with 0.5° Gaussian smoothing - for each (A-F) of the six alternative models for frequency-magnitude distribution in Southern California (seismic events from the SCEC catalog with M ≥ 2.7 on a grid of 12x12 cells are used - which guarantees completeness; data from 1/1/1981 to 31/12/2009 are employed for calibration, while from 1/1/2010 to 31/12/2025 for testing).

## Authors contribution

D.Z.: Conceptualization, methodology, validation, software, analysis, and visualization, writing-first draft, review and editing.

D.S.: Conceptualization, methodology, visualization, project administration, supervision, funding acquisition, writing-first draft, review and editing.

During the final stage of preparing this manuscript, the authors used DeepSeek and ChatGPT to improve the clarity and fluency of certain sections. All AI-generated suggestions were carefully reviewed and edited by the authors to ensure accuracy and consistency with the original work. The authors confirm that the content is original and take full responsibility for the published article. All authors agreed with the final version of the manuscript.

## Declaration of competing interests

Authors declare no competing interests.

## Fundings

This work is partially supported by the National Natural Science Foundation of China (Grant no. T2350710802, U2039202), and the Center for Computational Science and Engineering at Southern University of Science and Technology.